\documentstyle[referee]{l-aa} 
\input psfig.tex

.tfm    
\def\ni{\noindent}

\def\deg{\ifmmode^\circ\else$^\circ$\fi}

\def\lsim{\lower.5ex\hbox{$\; \buildrel < \over \sim \;$}}
\def\gsim{\lower.5ex\hbox{$\; \buildrel > \over \sim \;$}}
\begin{document}
\title {Ejection of the inner accretion disk in GRS 1915+105: the magnetic rubber-band effect}
\author{A. Nandi$^1$, S. K. Chakrabarti$^{1,2}$, S. V. Vadawale$^3$ and A. R. Rao$^3$}
\institute{ $^1$S.N. Bose National Center for Basic Sciences, Salt Lake, Kolkata 700 098, India\\
$^2$ Centre for Space Physics, 114/v/1A Raja S.C. Mullick Rd., Kolkata, 700047, India\\
$^3$Tata Institute of Fundamental Research, Homi Bhabha Road, Mumbai(Bombay) 400 005, India\\}
\offprints {A. Nandi     ({\it anuj@boson.bose.res.in })}
\date{Received 19 April 2001 / Accepted 4 October 2001; ASTRONOMY AND ASTROPHYSICS v. 380, p. 245, 2001}
\thesaurus{02.01.2; 02.02.1; 08.23.3; 08.9.2 GRS 1915+105; 13.25.5}
\maketitle
\markboth{Nandi et al: Ejection of inner accretion disk in  GRS~1915+105}{}

\maketitle

\begin{abstract}
We examine theoretically the behaviour of the inner accretion disk in GRS~1915+105
when soft X-ray dips are present in the X-ray light curve. We 
assume the presence of a radial shock in the accretion disk, as
in some of the Two Component Advective Flow (TCAF) solutions. We discuss the 
behaviour of the flux tubes inside a TCAF (which we name Magnetized
TCAF or MTCAF model for brevity) and compare various competing forces on the flux tubes.
In this MTCAF model, we find that the magnetic tension is the strongest force 
in a hot plasma of temperature $\gsim 10^{10}$K and as a result,
magnetic flux tubes entering in this region collapse catastrophically,
thereby occasionally evacuating the inner disk. We postulate 
that this magnetic `rubber-band' effect induced evacuated disk matter
produces the blobby components of outflows and IR/radio jets. 
We derive the size of the post-shock region by 
equating the time scale of the Quasi-Periodic Oscillations 
to the infall time of accreting matter in the post-shock 
region and found the shock location to be $\sim 45-66 r_g$.
We calculate the transition radius $r_{tr}$, where the Keplerian disk deviates
into a sub-Keplerian flow, to be $\sim 320r_g$.
Based on the derived X-ray spectral parameters,
we calculate the mass of this region to be $\sim$10$^{18}$g.
We conclude that during the X-ray dips the matter in the post-shock region,
which manifests itself as the thermal-Compton component in the
X-ray spectrum, is ejected, along with some sub-Keplerian matter
in the pre-shock region. 


\end{abstract}

\medskip

\section{Introduction}

GRS~1915+105 has proven to be an ideal source to study
in detail many of the physical concepts regarding accretion onto
black holes. Ever since its discovery (Castro-Tirado et al.
1992), it has been continuously bright in X-rays
and it shows a variety of X-ray variability characteristics
(Morgan, Remillard and  Greiner 1997; Muno et al. 1999; 
Yadav et al. 1999; Belloni et al. 2000). It has been 
monitored extensively in the radio band (Mirabel and 
Rodriguez 1994; Pooley and Fender 1997; Fender et al. 1999)
and several episodes of high radio emissions,
huge flares associated with superluminal motions, radio 
oscillations etc. are observed in this source.
Several attempts have been made to associate 
the radio emission, presumably coming from
jets, to the X-ray emission from the accretion disks (Fender et al.
1999; Naik et al. 2000; Naik and Rao 2000). The Spectral signature of 
winds from the Comptonising region has also been identified
(Chakrabarti et al. 2001).

Chakrabarti and Manickam (2000, hereafter CM00) have 
applied the Two Component Advective Flow (TCAF) model of 
Chakrabarti and  Titarchuk (1995) to explain 
various X-ray variability characteristics of 
GRS~1915+105. Recently there has been overwhelming evidence that
the TCAF model is valid for many black hole candidates 
(Smith, Heindl and Swank, 2001; Smith et al. 2001). CM00 invoked outflows from 
the inner accretion disk to explain a correlation between the
QPO frequency and the duration of the burst-off states during a
regular oscillations seen in the source. These outflows,
however, are confined to the sonic sphere and fall back on the
accretion disk after being cooled down by an inverse Compton effect. 
It was pointed out by Naik and Rao (2000) that the
source does not show appreciable radio emission during such 
oscillations. A detailed inflow/outflow model has not been
presented for this source to explain the radio emission,
particularly to explain the superluminally moving radio blobs.
 
Recently Naik et al. (2000) have detected a series of soft X-ray
dips during the declining phase of a huge radio flare and have
postulated that such soft dips are responsible for the 
jet emission. Vadawale et al. (2001) made a detailed study
of X-ray dips observed during the radio flare using the
Rossi X-ray Timing Experiment (RXTE) data and have presented 
evidence for the disappearance of the inner accretion disk 
during the dips. Since the disappearance of the inner disk is 
seen to be correlated with intense radio activity, 
the role of the magnetic field must be studied in order to 
understand the system completely. Rodriguez and Mirabel 
(1999) estimated the field in radio blobs to be
around tens of mG at 500-1000AU (in 1994 observation). 
Fender et al. (1997) requires the field to be 
around 8G at around 1AU (in their 1996 observations). 
From the similarity of $\sim 30$min oscillations in IR and Radio,
they concluded that the radio blobs are adiabatically expanding
and are independently ejected from the disk every $30-40$ minutes.
If the trapped field inside a radio blob is of roughly $1/r$ (for toroidal
field) then its interpolated value close to a black hole
is around $10^{7}$G at around $10r_g$ which is 
comparable to an equipartition  value. Thus, one needs to correlate 
fields ejected from the disk with those observed inside the 
radio blobs.  In this Paper, we examine the mass ejection
based on the TCAF model in presence of a magnetic field 
(we call this as Magnetized TCAF or MTCAF model)
amplified due to strong shear at the transition radius of 
the Keplerian and a sub-Keplerian flow. In the next Section,
we discuss forces which govern the motions of the flux tubes
and write equations of toroidal flux tubes inside 
an accretion disk with a constant angular momentum.
We show that close to the black hole, where the flow could be 
very hot ($\gsim 10^{10}$K)the flux 
tubes move at least with the Alf\'ven speed and may
catastrophically collapse like a stretched rubber band. 
We conjecture that such rapid collapse would assist 
evacuation of matter from the disk and cause X-ray `dips' 
seen in the light curves.  In \S3 we estimate the mass of the
ejecta which agrees with observations. Finally, in \S4 we draw 
our conclusions.

\section{The Magnetized TCAF model for GRS~1915+105}

Based on the global solutions of the most general advective
accretion disk solution (Chakrabarti, 1990, 1996a), 
Chakrabarti and Titarchuk (1995) presented a TCAF model 
of accretion onto black holes. According to this model, 
matter with high viscosity and angular momentum settles
into the equatorial plane, while matter with lower 
viscosity and angular momentum surrounds the Keplerian 
disk. This sub-Keplerian component is likely to form a 
standing or an oscillating shock (Molteni, Sponholz and 
Chakrabarti, 1996; Ryu, Chakrabarti and Molteni, 1997) 
front near the centrifugal barrier (few tens of Schwarzschild 
radii) depending on whether the Rankine-Hugoniot condition is
satisfied or the cooling time in the post-shock region is 
comparable with the infall time. Soft photons from the 
Keplerian disk in the pre-shock flow are intercepted 
by the puffed up sub-Keplerian post-shock flow and are 
reprocessed due to inverse Comptonization. If the post-shock 
matter remains hot, the black hole exhibits a harder spectrum, 
while if the post-shock region is cooled down by the
soft photons, the black hole exhibits a soft spectrum. There
could be a sub-Keplerian region just before the shock 
as well since with low viscosity and accretion rates a
Keplerian disk recedes from a black hole and it is not 
necessary that the shocks form right at the region 
where a transition from Keplerian to sub-Keplerian flow takes place.
Fig. 1 schematically shows this behaviour of the two components.

The centrifugal pressure supported boundary layer (or, CENBOL for short)
formed in a transonic, advective flow (Chakrabarti et al. 1996),
has most of the features of a thick
accretion disk although in advective disks, advection is included 
self-consistently and in thick disks advection is totally ignored. 
This is because at the CENBOL surface, matter undergoes a supersonic 
to sub-sonic transition and it moves very slowly in the radial
direction. In some phases of accretion, matter can bring in a large 
stochastic magnetic field. The field is sheared 
due to strong azimuthal velocity and the toroidal field
becomes very strong. These field lines will have very little
matter within it and would likely be buoyant and emerge 
from various parts of the accretion disk. Fig. 1
schematically shows this behaviour. Toroidal magnetic 
flux tubes released from the Keplerian disk are sheared, 
amplified and are advected in the sub-Keplerian flow. 
Due to the centrifugal barrier matter stays away from the axis. 
Thus a so-called funnel wall is created exactly as in 
a thick accretion disk (Paczy\'nski \& Wiita, 1980).
Chakrabarti and D'Silva (1994, hereafter CD94) computed the 
nature of their trajectories inside a thick accretion disk. 
They showed that in the event a strong flux tube enters a
hot region with ion temperature $T_i \gsim 10^{10}$K, the 
magnetic tension becomes the strongest force and the flux 
tube catastrophically collapses. Because of strong similarity 
of the thick accretion disk and an advective flow, especially 
inside the CENBOL, we believe that a similar mechanism could be 
working and flux tube collapse would take place. 

We shall consider the motion of the flux tubes on the equatorial plane of an
accretion flow around a Schwarzschild black hole described by 
Paczy\'nski-Wiita (1980) pseudo-Newtonian acceleration $g=-1/2(x-1)^{-2}$.
We use the geometric units. Masses are measured in units of the 
mass of the central black hole, $M_{BH}$; distances from the axis ($x$) will be 
measured in units of the Schwarzschild radius $r_g=2GM_{BH}/c^2$;
and the time scales are measured in units of $r_g/c$. Inside the 
disk, we choose the polytropic equation of state, $P=K\rho^\gamma$, 
where $K$ and $\gamma$ are constants. The magnetic flux tubes brought 
in by advection are assumed to be sheared and {\it axisymmetric} toroidal 
flux tubes of random shape and size could be produced inside the flow. We 
however assume that the flux tubes are thin, i.e., the flux tube 
cross-sectional radius $\sigma$ is smaller than the local pressure 
scale height of the disk. Close to a black hole, angular momentum of the flow remains constant 
(Chakrabarti, 1996a) even in presence of moderate viscosity.
Thus, we choose specific angular momentum $\lambda$ in the 
sub-Keplerian region to be constant.
The equations of motion for thin flux tubes 
have been given in CD94 and we do not repeat here.
For the sake of completeness, we write down the radial equation only valid for 
the equatorial plane ($\theta=\pi/2$),
$$
\ddot{x} + {X\over{(1+X)}}[-x\dot{\phi}^2 - 2x\omega\dot{\phi}] 
= {X\over{(1+X)}}\{{M_b\over X}[g - x\omega^2] -
\frac{1}{m_i}\frac{\psi^2}{2\pi\sigma^2} - {D_x\over{\pi\sigma^2\rho_e}}\}, 
\eqno{(1)}
$$
\ni where $X=m_i/m_e$, $m_i = 2 \pi^2 \sigma^2 x \rho_i $ and
$m_e = 2 \pi^2 \sigma^2 x \rho_e$ are the masses of
the fluid inside and the fluid displaced  by the flux tube respectively,
$\rho_i$ and $\rho_e$ being the corresponding densities.
Subscripts $e$ and $i$ indicate whether the relevant quantity is of the ambient
(external) medium or within the flux tube (internal medium). $\psi=\pi \sigma^2 B$,
$B$ being magnetic field of the tube and $\dot \phi$ is the intrinsic angular velocity
of the flux tube inside the disk. The drag term is assumed to be similar to the 
drag experienced by a cylinder moving perpendicular to its axis in a fluid, and the drag per 
unit length in radial direction is given by,
$$
D_x = -{1\over 2}C_D \rho_e\sigma(\dot{x}-u)^2
\eqno{(2)}
$$
where $C_D\sim 0.4$ (Goldstein 1938).
The pre-shock, sub-Keplerian flow is assumed to have a radial velocity, 
$$
u \sim  \beta/\sqrt{{(x-1)}}
\eqno{(3a)}
$$
and the post-shock sub-Keplerian flow,
$$
u \sim  1/R\sqrt{{(x-1)}}.
\eqno{(3b)}
$$
Here, $\beta$ is a factor by which sub-Keplerian matter slows down compared to
a freely falling flow. $\beta\sim 1$ for very low angular momentum cool flow. 
$R$ is the compression ratio of the shock by which matter is assumed to be 
slowed down inside the CENBOL. 

From Eq. (1), we note that there are four forces in 
operation: (i) The second term inside the bracket of the 
left hand side is the Coriolis force $F_c=2 v_\phi \omega$, where $\omega=\lambda/x^2$
is the angular velocity of the flow. The Coriolis acceleration is given by (CD94),
$$
a_C=\frac{\rho_i}{\rho_i+\rho_e}\frac{2\lambda^2}{x} [\frac{1}{x_0^2}-\frac{1}{x^2}].
\eqno{(4)}
$$
A flux tube brought from $x_0$ to $x$ inside a flow of constant angular 
momentum would feel no Coriolis force if the motion is along the direction of 
constant angular velocity since $x_0\sim x$ for rotating bodies.
A magnetic flux tube with buoyancy factor $M_b (=\frac{\rho_e-\rho_i}{\rho_e})=1-X$,
will feel the (ii) Magnetic Buoyancy force (first term on the right hand side 
inside the curly bracket). The corresponding acceleration is (CD94):
$$
a_{MB}=\frac{M_b}{1+X}[\frac{1}{2(x-1)^2} - \frac{\lambda^2}{x^3}].
\eqno{(5)}
$$ 
The tube will emerge out of the disk if $a_{MB}$ dominates over $a_C$.
The  middle term on the right hand side of Eq. (1) is the  (iii) force due to Magnetic Tension. The
corresponding acceleration  is $a_T$ (CD94):
$$
a_T=-\frac{1}{m_i+m_e} \frac{\Psi^2}{2 \pi \sigma^2} =-\frac{B^2}{4 \pi x (\rho_i+\rho_e)}.
\eqno{(6)}
$$
The final and very important force is represented by the final term of Eq. (1). It is
called the (iv) Drag Force exerted due to the motion of a rigid tube inside a 
flow. The corresponding acceleration is (CD94):
$$
a_{D}=-\frac{1}{2}\frac{C_D \rho_i \sigma ({\dot x}-u)^2 }{2 \pi \sigma^2 (\rho_i+\rho_e)} 
\sim -\frac{C_D \rho_i u^2}{2\pi\sigma (\rho_i+\rho_e)}.
\eqno{(7)}
$$
Here we used ${\dot x} \sim 0$ for computing the maximum value of the drag force.

%Assuming that initially, at the time of release, the flux tube is in thermal equilibrium
%with the surroundings, we have,
%$$
%p_{g,i}+\frac{B^2}{8\pi}= p_{g,e}.
%\eqno{(8)}
%$$
%Here, subscript $g$ denotes pressure due to gas.  Initial magnetic buoyancy could be written as the ratio 
%of the magnetic pressure to the total gas pressure, if the tube is in thermal equilibrium (CD94),
%$$
%M_b=\frac{B^2}{8 \pi p_{g,e}}.
%\eqno{(9)}
%$$
Equating the acceleration due to buoyancy with that due to magnetic tension (Eq. 6)
on the equatorial plane, we get the critical temperature of the external gas as (CD94),
$$
T_{p,0}=\frac{m_p c^2[\lambda_K^2(x_0)-\lambda^2]}{8 k x_0^2},
\eqno{(8)}
$$
above which the tension dominates over buoyancy and the flux tubes rush towards the funnel
wall catastrophically. Here we used the mean electron number per ion
to be $\mu=1/2$ and $k$ is the Boltzmann constant. Subscript $0$ specifically indicates that 
the flux tube will behave like an over stretched rubber-band only after it crosses 
$x=x_0$, where  $T>T_{p,0}$. Note that since we are dealing with a sub-Keplerian disk,
$T_{p,0}>0$ in the entire region of interest. For typical values $x_0=100$, $\lambda_0=1.8$,
one obtains $T_{p,0}\sim 6 \times 10^9$K. For an adiabatic disk, this assumption remains valid
even when the flux tube collapses very rapidly since its internal temperature will
increase adiabatically in the same way as in the external disk. For an efficiently cooled
two-temperature flow, the above proton temperature would correspond to an electron temperature of
$T_e \sim \sqrt{{m_e}/{m_p}} T_p$. This is around $14$keV which is very reasonable for the
temperature of the sub-Keplerian region. When the flux tubes fall radially, neither Coriolis force
nor the drag term could be neglected. In fact, accelerated flux tubes 
would have high ${\dot x}$ in the drag term as they move faster than the bulk radial motion.
Meanwhile, assuming that internal mass of a flux tube is roughly constant, the cross-section
$\sigma \sim 1/\sqrt{x}$ increases as the flux tube approaches the black hole. So, this will also
increase the drag term. As a result, we expect that the flux tube
would slow down somewhere close to the black hole and buoyancy
would eject the flux tube out of the disk perpendicularly as shown in Fig. 1. Typical trajectories of the
flux tubes, based on numerical integrations in CD94, are shown in Fig. 1.

\begin {figure}
\vbox{
\vskip 0.0cm
\centerline{
\psfig{figure=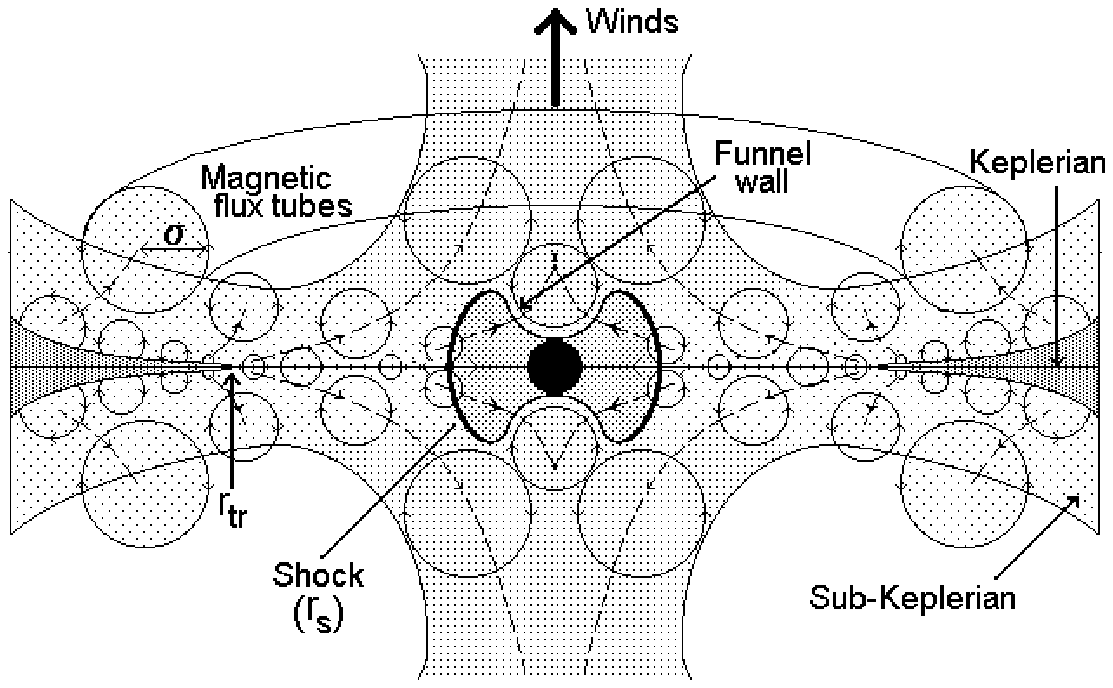,height=10truecm,width=10truecm}}}
\end{figure}
\begin{figure}
\vspace{-3.0cm}
\caption[] {A cartoon diagram of the accretion disk near a black hole
which includes a shock ($r_s$), a sub-Keplerian and a Keplerian disk with a boundary at $r_{tr}$.
Stochastic magnetic fields are sheared and amplified as they leave a Keplerian disk.
In a hot, sub-Keplerian flow, these toroidal flux tubes catastrophically collapse
squirting matter along the axis, and thereby evacuating the disk and 
producing outflows. Dashed curves show typical trajectories (CD94).}
\end{figure}

The buoyancy timescale $t_b$ is estimated from Eq. (5): 
$a_{MB} \sim h/t_b^2 = \frac{M_b}{(1+X)x^3} (\lambda_K^2-\lambda^2)$, 
where, $h \sim a_s x^{3/2}$ is the half thickness of the disk in vertical 
equilibrium at $x$ and $a_s \sim x^{-1/2}$ is the speed of sound. 
Note that $t_b$ is very large in a Keplerian disk ($\lambda_K \sim \lambda$).
For a sub-Keplerian flow, $\lambda/\lambda_K \sim 0.4$ with $\lambda=1.8$ at $x=30$,
$X\sim 0.1,\ M_b \sim 0.9$, $t_b\sim 240$.

Collapse time $t_f$ of a flux tube is estimated in the following way:
When tension is the most dominant force, the radial 
equation of the flux tube (Eq. 1) is simplified to:
$$
{\ddot x} + a_T=0,
\eqno{(9)}
$$
Putting ${\ddot x} \sim x/t_f^2$, we see 
that the velocity of collapse of the flux tube $v_f$ is,
$$
v_f \sim (\frac{B^2}{4\pi \rho_e})^{1/2} \sim v_a 
\eqno{(10)}
$$
where $v_a$ is the Alf\'ven velocity. Thus, flux tube collapses in Alf\'ven speed.
Since $B \propto x^{-2}$, magnetic pressure is $\propto x^{-4}$,
while the gas pressure $p_e \propto x^{-5/2}$, 
the ratio of magnetic to gas pressure $f \propto x^{-3/2}$, increasing with 
the decrease of $x$. As the flux tube leaves the Keplerian disk, large shear 
at the transition region and in the sub-Keplerian flow quickly 
intensifies the flux tube to $M_b \sim 1$ during infall. On the other hand, if
$f \sim 1$ at $x\sim 300$ where the flow deviates from a Keplerian disk,
$f \sim 30$ at $x=30$, inside the CENBOL. The Alf\'ven speed $v_a \sim
(\frac{B^2}{4\pi\rho_e})^{1/2} \sim (f/\gamma)^{1/2} a_s \sim 5 a_s$
where we used $\gamma=4/3$.  Since, inside a CENBOL, the velocity of matter $v_m$ is highly sub-sonic,
it is therefore high sub-Alf\'venic:
$v_m <<v_a$. Thus, $v_f/v_m >>1$ and the flux tube
collapses catastrophically. This justifies
the conjecture made earlier that the rubber-band effect 
could evacuate the disk (Chakrabarti, 1994; 1996b; 2000). For 
$a_s \sim x^{-1/2}$, $t_f \sim x^{3/2}/5 \sim 30$
at $x=30$. Thus inside the CENBOL $t_b>>t_f$.

So far, we have ignored many non-axisymmetric effects
such as Parker instability (1979 and references therein) and shearing 
instability (e.g., Balbus \& Hawley, 1991). Foglizzo \& Tagger (1995) treated this problem
comprehensively in the context of a standard disk embedded in a 
large scale field. They found that (a) if the wave-length is larger than the
disk thickness then the flux tube is very unstable and buoyantly comes
out of the disk and (b) instability is strongest if the field is weaker.
According to Parker (1979), submerged field tubes may break
up into filaments in timescales of around $3 \Lambda/v_a$, where $\Lambda$
is the scale height and the field is able to escape from the gas
in timescales $t_P \sim \Lambda/v_a$. If $\Lambda \sim h \sim x$, which is especially true
in CENBOL region, the time scale of the
escape of the field may be comparable to the $t_f$ as obtained above.
As a result, the flux not only collapses
catastrophically, but also escapes upwards following a curved trajectory as depicted in Fig. 1.
In presence of a differential rotation, Balbus and Hawley (1992) suggested that even a small 
initially vertical field would be amplified to create all the components
in the dynamical timescale $t_d \sim 1/\omega = x^2/\lambda \sim 50$ (at $x=30$, $\lambda=1.8$). 
Numerical simulation (Hawley, Gammie \& Balbus, 1995) has verified this instability
neglecting the tension effects. Even though the time scales of the collapse, the Parker 
instability and the shear instability are of the same order, we believe that the
rubber-band effect would still be important for the destruction of the inner disk.

\section{Estimation of mass of the post-shock region and the sub-Keplerian region}

Vadawale et al. (2001) have shown that during the soft X-ray dips
a thermal-Compton component in the X-ray spectrum gets suppressed.
Several works in the literature talk about the disk-evacuation
(e.g., Belloni et al. 1997; Feroci et al. 1999) in this context.  We like to 
understand this using Two Component Advective Flow (TCAF) models
of Chakrabarti \& Titarchuk (1995) in presence of a Magnetic field (i.e., MTCAF model) 
and its time variability properties described in Molteni, Sponholz \& Chakrabarti (1996); Ryu, 
Chakrabarti \& Molteni (1997) and Chakrabarti \& Manickam (2000). 
The observation of possible disk-evacuation is clearly in line with the
TCAF model and the shock oscillation model of the quasi-periodic oscillations
(CM00; see also, Rao et al. 2000) which showed that the Comptonising post-shock 
region participates in oscillation. The sub-Keplerian region in the pre-shock 
flow does not emit much radiation and it is possible that some of this region 
may also be disrupted during the rapid collapse of the flux tube. 
Once we accept the destruction of the sub-Keplerian region by the magnetic
rubber-band effect, we can compute the mass of this region in the following way:

The shock location is computed by equating the infall time from the shock with the
time scale of QPO. This time scale $t_{ff}$ can be written in the form (CM00):
$$
t_{ff}^{-1} = \frac {1}{R} \frac{1}{r_s^\alpha} \frac{c v_0}{r_g},
\eqno{(11)}
$$
where, $R$ is the compression ratio (see also, Eq. 3b), $r_s$ is the shock location, 
$v_0$ is a dimensionless constant. Here, $\alpha =3/2$ for free-fall
motion and $\alpha=1$ for a flow of constant velocity $cv_0/R$ 
in the post-shock region. Using this assumption, the shock 
location in the pre-dip and the post-dip flow (for parameters in 
these flow see, Vadawale et al. 2001) are $\sim 45r_g$ and 
$\sim 66 r_g$ respectively. CM00 proposes that a better 
fit in the correlation between the duration of the QPO and the frequency 
of QPO requires more or less constant velocity in the post-shock region
with a rough velocity of $0.066c/R$. This produces the shock
in the pre-dip and the post-dip  flow at $79r_g$ and $140r_g$ respectively.
Using the parameters of Vadawale et al. (2001), and the location of the
shock as given above, the electron number density  
in the Comptonising region is found to be $n_e \sim 10^{17}$ cm$^{-3}$. The 
corresponding mass ($\frac{4}{3}\pi r_s^3 m_p n_e r_g^3$) 
of the region is $M_{CENBOL} \sim 2-3 \times 10^{18}$g depending on the model of the inflow.
How much of this matter is squirted out of the disk along the axis?
Again, in the absence of the size distribution of the flux tubes,
the answer is difficult. However, because of centrifugal force, matter is 
unlikely to enter within the `funnel wall' (CD94) even after compression,
sudden collapse of large flux tubes [$\sigma \sim h(r_s)$]
would  be expected to displace the whole CENBOL region parallel to the 
funnel wall. The outflow rate would be ${\dot M}_{out} \sim M_{CENBOL} /t_f$
which may be very large compared to the inflow rate for a short duration of $t_f$.
If flux tubes are smaller in size, since the collapse velocity is
much larger than the speed of sound, matter will still be displaced 
but would be refilled in free-fall time, unless there are many near-simultaneous flux
collapse events. 

If the magnetized sub-Keplerian disk is removed by imploding flux tubes 
as described in the earlier Section, one requires to know the location of the inner edge of 
the Keplerian disk to estimate the complete mass involved. 
From the model fit (Vadawale et al. 2001), the Keplerian disk 
temperature turns out to be $T_K \sim 1.5$keV. With a hardening factor 
of around $1.7$ (Shimura and Takahara, 1995), the mass of the black hole as $10 M_\odot$ 
and Shakura-Sunyaev viscosity parameter $\alpha_{SS} =0.01$, the above 
temperature corresponds to a transition radius at around $r_{tr} \sim 320r_g$ 
(Shakura and Sunyaev, 1973). Assuming density falling off as $\rho \sim \rho_0 
(\frac{x_0}{x})^{3/2}$, the mass of the sub-Keplerian flow of size $x_{tr}$ 
($\int 4\pi x h(x) \rho dx$ with $h(x) \sim x$) is around $10^{20}$g. 
These computations assume no pair production, i.e., there is exactly one 
electron for each proton in the Comptonising region.

Once the evacuation is complete, the disk is filled in quickly 
by sub-Keplerian matter in timescale of:
$$
t_{visc} \sim \frac{1}{\alpha_{SS}}(\frac{h(x)}{x})^{-2}\frac{x}
{v_{Kep}}=192(\frac{0.01}{\alpha_{SS}}) (\frac{0.03}{a_s})^2
(\frac{x_{tr}}{300})^{1/2} \frac{M_{BH}}{10M_{\odot}}{\rm s}.
\eqno{(12)}
$$
Here, $h(x) \sim a_s x^{3/2}$ ($a_s$ is the speed of sound)
is the local vertical height of the disk. The time scales seem
to be reasonable, since the dips have been seen to be filled up in a matter of $150$ 
to $200$ seconds. It is to be noted that Dhawan et al. (2000) using 
disk-instability model of Belloni et al. (1997) obtained the missing inner disk region to be
only $180$km. This distance is about $6r_g$ and with $10^{18}$gm s$^{-1}$ accretion rate
which they employ, the mass of this region cannot be enough to create the $10^{23}$gm
blob (or, even $10^{18}$g mini-blobs) observed by Mirabel \& Rodrigu\'ez (1999).
This is particularly because the region $1-3r_g$ is definitely supersonic and sub-Keplerian 
and therefore  is of very little mass. We believe that the missing region should be much 
larger, possibly order of a $100~r_g$ or so.

\section{Conclusion}

In this Paper, we have given a physical basis for a sudden
mass ejection in GRS~1915+105. We showed that if matter brings in a particularly 
strong magnetic field, this would be sheared and amplified to a value much above the
equipartition value before it can be expelled by buoyancy. Magnetic tension 
collapses these toroidal flux tubes at a highly supersonic speed, much faster 
than the flow velocity. This has the effect of displacing matter from the
disk in transverse direction (much like a fast boat causing spillage
on a shore) and depositing it to outflowing winds. From the observed fits of 
Vadawale et al. (2001) we estimated the electron number density and the mass of the
post-shock region and the sub-Keplerian flow to be around $10^{18}$g 
and $10^{20}$g respectively. Our estimate of the post-shock mass
is a factor of ten less than the mass estimate ($10^{19}$g) 
of `baby-jets' (Mirabel et al 1998) associated with IR and radio flares
and could therefore be due to ejection of some sub-Keplerian matter as well.
These `baby-jets' are found to be associated with class $\beta$ light curves 
which have soft X-ray dips. These dips are also seen in class $\theta$ light curves.
During a major portion of the huge radio flares associated with superluminal blob
emission a series of soft dips are present (Naik et al. 2000).
Mirabel and Rodriguez (1999) have pointed out that in each epoch of this type of outflow, 
the mass condensation is around $10^{23}$g. In order to achieve this, we
require that matter is accumulated from disk evacuation at each `dip' and 
within each epoch, successive mini-blobs move faster than the earlier blob
in order to `catch up'. This may indicate some other runaway process with a 
longer time scale of tens of days. Naik et al (2000) have observed such X-ray dips 
at a rate of once in a few hundred seconds during the peak or the radio 
flare. If there are $\sim 1000$ evacuation events during an episode of superluminal blob
ejection (in a few days), then total mass condensation would be $10^{23}$g.
Hence in order to explain the observation of Mirabel and Rodriguez (1999),
one must require that in each epoch, matter is accumulated from at least 
a thousand evacuation events. Future observation would tell if such is the case.

It is to be noted that the mass of the condensation as estimated by Mirabel
and Rodriguez (1999) is based on the presence of one electron per proton, i.e.,
no pair production is assumed. With a pair density, say, ten times larger,
the estimated mass would be ten times less. However, at the same time, estimated mass
of the disk would also be reduced by the same factor. Hence, the number of ejection
events is not affected. 

\begin{acknowledgements}
SKC and AN acknowledge receiving grants from DST project
entitled Analytical and Numerical Studies of Astrophysical Flows Around Compact Objects.
The authors thank the referee for helpful comments.
\end{acknowledgements}

{}

\end{document}